\documentclass[12pt]{article}
\usepackage[usenames,dvips]{pstcol}
\usepackage{color}
\usepackage{amssymb}
\usepackage{epsfig}
\usepackage{footnote}
\usepackage{longtable}
\usepackage{verbatim}
\usepackage{array,multirow}
\usepackage{titlesec}
\usepackage{graphicx}
\usepackage{caption}
\usepackage{epstopdf}
\def\LAPPDTM{LAPPD\textsuperscript{TM}~}
\def\degreesC{$^o$C~}
\def\K2CsSb{$K_2CsSb$~}
\def\Cs3Sb{$Cs_3Sb$~}
\def\Al2O3{$Al_2O_3$~}
\newrgbcolor{maroon}{.45 0.1 0.1}
\begin{document}
\pagestyle{plain}
%

%
%
\begin{center}
{\Large\bf Air-Transfer Production Method for Large-Area Picosecond
Photodetectors}
\end{center}

\begin{center}
E. Angelico, A. Elagin, H. J. Frisch, E. Spieglan\\
{\it Enrico Fermi Institute, the University of Chicago}\\

B. W. Adams\footnote{Current Address: Quantum Optics Applied
  Research, Naperville IL}, M. R. Foley, M. J. Minot\\
{\it  Incom, Inc}\\

\today\\
\end{center}

\begin{abstract}
 We have designed and prototyped the process steps for the batch
 production of large-area micro-channel-plate photomultipliers
 (MCP-PMT) using the ``air-transfer'' assembly process developed with
 single $\LAPPDTM$ modules. Results are presented addressing the
 challenges of designing a robust package that can transmit large
 numbers of electrical signals for pad or strip readout from inside
 the vacuum tube and hermetically sealing the large-perimeter
 window-body interface. We have also synthesized a photocathode in a
 large-area low-aspect-ratio volume, and shown that the micro-channel
 plates recover their functionality after cathode synthesis. The
 steps inform a design for a multi-module batch facility employing
 dual nested low-vacuum (LV) and ultra-high-vacuum (UHV) systems in a
 small-footprint. The facility design provides full access to multiple
 MCP-PMT modules prior to hermetic pinch-off for leak-checking and
 real-time photocathode optimization.
\end{abstract}

\setcounter{tocdepth}{3}
%
%
\section{Introduction}
\label{introduction}
Applications for large-area coverage of photodetectors with excellent
time and space resolution include: correlated precision time and space
information in high-energy particle collider and fixed-target
events~\cite{LAPPD_proposal}; simultaneous measurements in near and
far detectors of neutrino oscillations using time-selected energy
spectra~\cite{stroboscopic_paper}; imaging double-beta decay events in
large liquid scintillator detectors using timing to separate Cherenkov
and scintillation
light~\cite{OTPC_paper,Andrey_paper_1,Andrey_paper_2,Erics_CPAD,Andrey_paper_3,Andrey_paper_4};
reconstruction of neutral mesons decaying to photons in searches for
rare meson decays~\cite{Philadelphia_2017_proceedings,KOTO,Redtop};
low-dose whole body Time-of-Flight Positron-Emission
Tomography~\cite{Heejong_NIM_2009,Heejong_NIM_2010,Heejong_NIM_2011,PET_patent};
and 6) nuclear non-proliferation and reactor
monitoring~\cite{Watchman}.

\begin{figure}[htb]
\centering
\includegraphics[angle=0,width=0.50\textwidth]{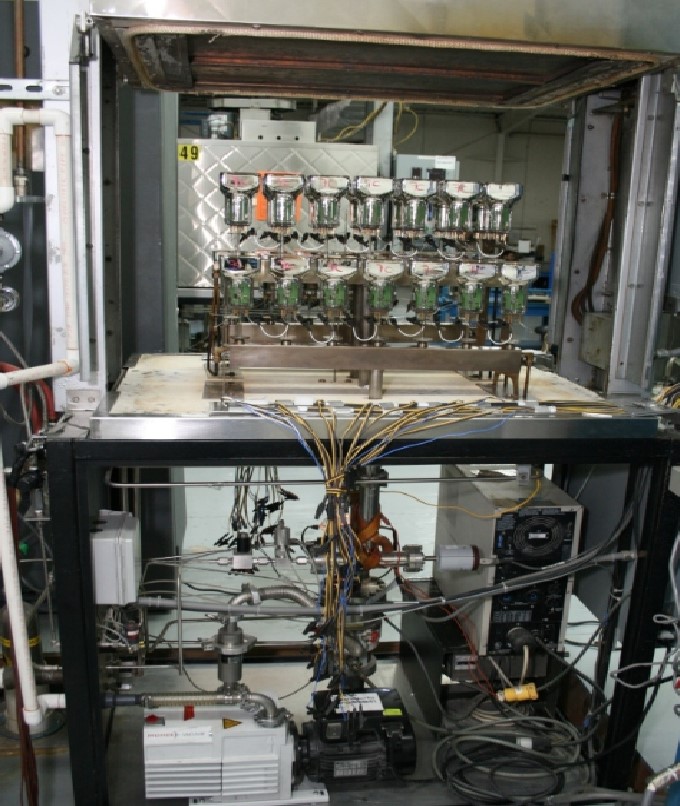}
\caption{The photomultiplier batch production station purchased from
  the Photonis Lancaster plant~\cite{Photonis} that served as the
  model for development of the LAPPD batch production process.}
\label{fig:burle_station}
\end{figure}

The typical method of production of commercial photomultipliers (PMT)
is in a batch production station that processes multiple modules per
cycle. Figure~\ref{fig:burle_station} shows a commercial PMT
station~\cite{Photonis} purchased by the LAPPD
Collaboration~\cite{history_paper}.  The PMT process steps served as
the model for the development of the batch production process for
large-area flat-panel MCP-based photodetectors (MCP-PMT)
~\cite{Batch_patent}.

The salient features that form the basis of the design for flat-panel
photodetector production are as follows~\cite{Photonis}. During
processing the PMT glass envelopes are at ultra-high vacuum (UHV)
internally and are at atmosphere externally. The initial evacuation of
the phototube is at low vacuum (LV), performed by an external pump
through long low-conductance plumbing, with the PMT glass envelopes
connected through a narrow tubulation sealed by a low-vacuum O-ring
gland.  An oven is placed over the assembly for bake-out and
subsequent photocathode formation.  The antimony pre-cursor layer is
evaporated by line-of-sight onto the inside surface of the entrance
window, and the alkali vapor is generated by sources inside the glass
envelope. The UHV inside the tube is provided by the bake-out, the
internal getter, and the alkali vapors, and not by the external pump.
After cathode optimization the tubulation is pinched off hermetically
sealing the phototube with the getter inside.

Experience with the process steps for the Photonis station guided the
development of the process for large-area micro-channel-plate-based
PMT batch production described below. However, large-area
MCP-PMTs present unique challenges, requiring
a robust package that can transmit large numbers of electrical signals
for pad or strip readout from inside the vacuum volume, hermetically
sealing the large-circumference window-body interface, synthesizing
the photocathode in a large flat package, and recovery of the
functionality of the large surface-area MCPs after cathode synthesis.

The synthesis of bialkali photocathodes by exposing an antimony (Sb)
layer pre-deposited on a substrate by thermal evaporation was
described by Sinclair in 2009~\cite{sinclair_suggestion}.  The
suggestion led to a DOE-funded development program for a theory-based
\K2CsSb cathode in a collaboration of Chicago, BNL and the RMD
corporation~\cite{RMD_SBIR_1,RMD_SBIR_2,Smedley_IPAC2015,
  Mengjia_cathode_paper}.  However, we recently discovered that the
chemistry of photocathode synthesis based on a pre-deposited Sb layer
had been explored in detail by Barois et al. in
1989~\cite{barois_paper_1,barois_paper_2}, and that the air-transfer
process is currently being used commercially in Russia to make
multi-anode photomultipliers~\cite{MELZ}.

The Barois insitu photocathode process tightly controls the uniformity
of the cathode layer through the thickness of the Sb precursor layer,
which can be made highly uniform by conventional thermal
evaporation~\cite{barois_quote}.  The chemical and physical
composition of the structure is determined by a number of parameters
as discussed in
Refs~\cite{RMD_SBIR_1,RMD_SBIR_2,Smedley_IPAC2015,Mengjia_cathode_paper},
and has the attractive property that the reaction runs to completion.
While it is possible to push the photocathode beyond its optimal
stoichiometry for quantum efficiency~\cite{barois_paper_2}, the
uniformity is governed largely by the initial antimony
layer~\cite{barois_quote}.

%
%
\section{The Dual-Vacuum Air-Transfer Method}
\label{dual_vacuum}

The LV/UHV-dual-vacuum technique uses an outer vacuum vessel
which contains the photodetector module. The outer vessel provides thermal
insulation for a high-temperature bakeout of the photodetector module 
(``tile''), reduces
the formation of oxides in the solder during heating, and ensures that no
mechanical pressure differential exists across the molten seal.  The
module is assembled with the entrance window clamped in place, and is connected
through a tubulation or tubulations to an independent UHV manifold and
pump.  Initially, before the hermetic seal is formed during the
bake-out thermal cycle, the two vacuum systems can communicate through
the unsealed seam between the window and the tile base.  The
bake-out cycle forms the seal between the window and base, and
activates the getter in the module. After cooling below the seal
melting temperature the two vacuum systems are independent, and the
outer vacuum can be opened to atmosphere.  The inner vacuum system is
used as a source for the alkali vapors to synthesize the
photocathode from the Sb precursor layer on the window.

\begin{figure}[htb]
\centering
\includegraphics[angle=0,width=0.50\textwidth]{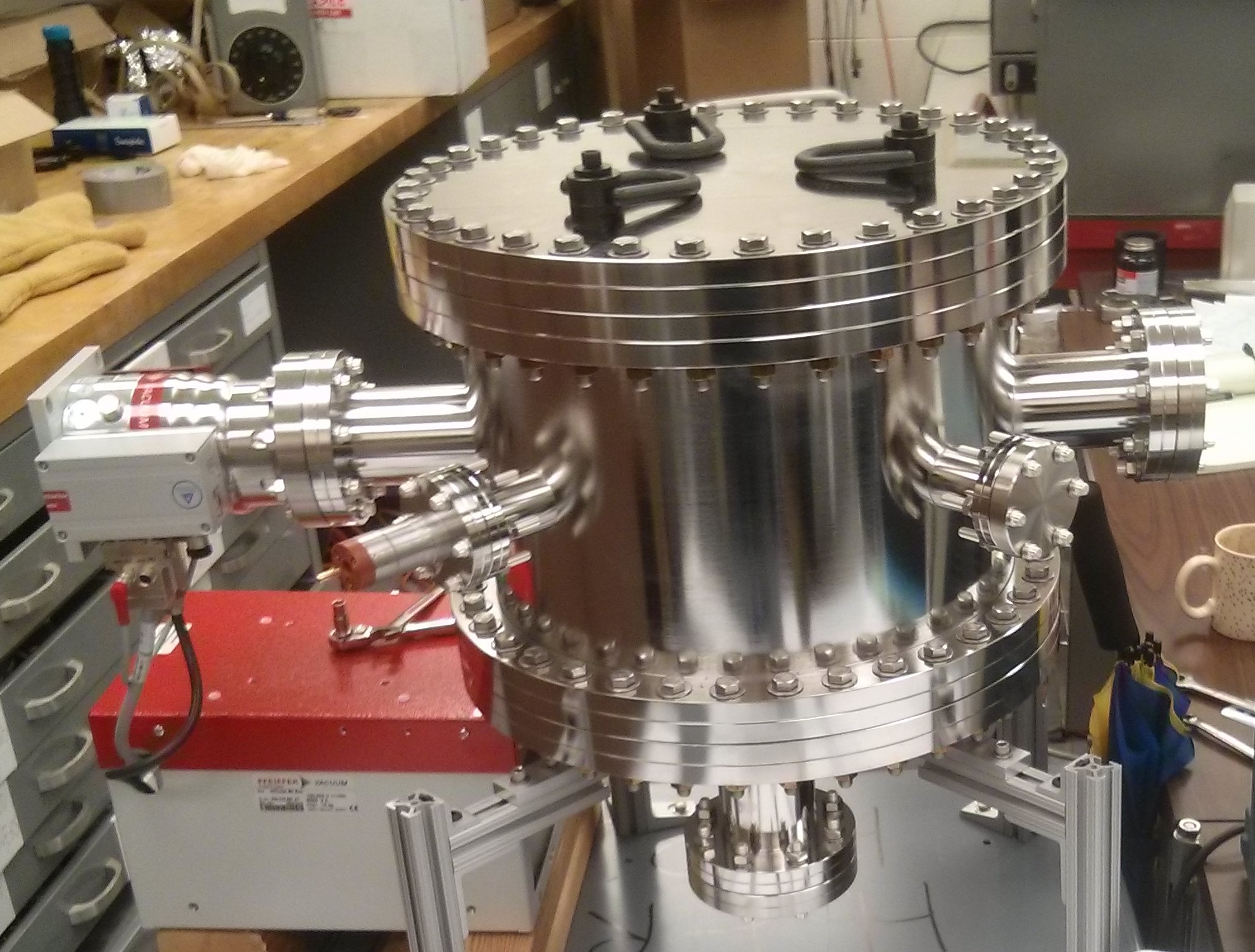} 
\caption{The outer LV vacuum vessel that provides thermal insulation
  to the photodetector module during bake-out, a low oxidation
  environment, and equalization of the pressure across the molten
  indium alloy seal. }
\label{fig:margheritaII}
\end{figure}

Figure~\ref{fig:margheritaII} shows the outer vacuum
vessel. The vacuum pump for the outer system is on
the left; the connections to the inner vacuum system enter through the
flange on the bottom. The upper part of the vessel can be lifted off
after sealing, exposing the sealed module for
leak-checking and subsequent
photocathode synthesis. 

\begin{figure}[htb]
\centering
\includegraphics[angle=0,width=0.50\textwidth]{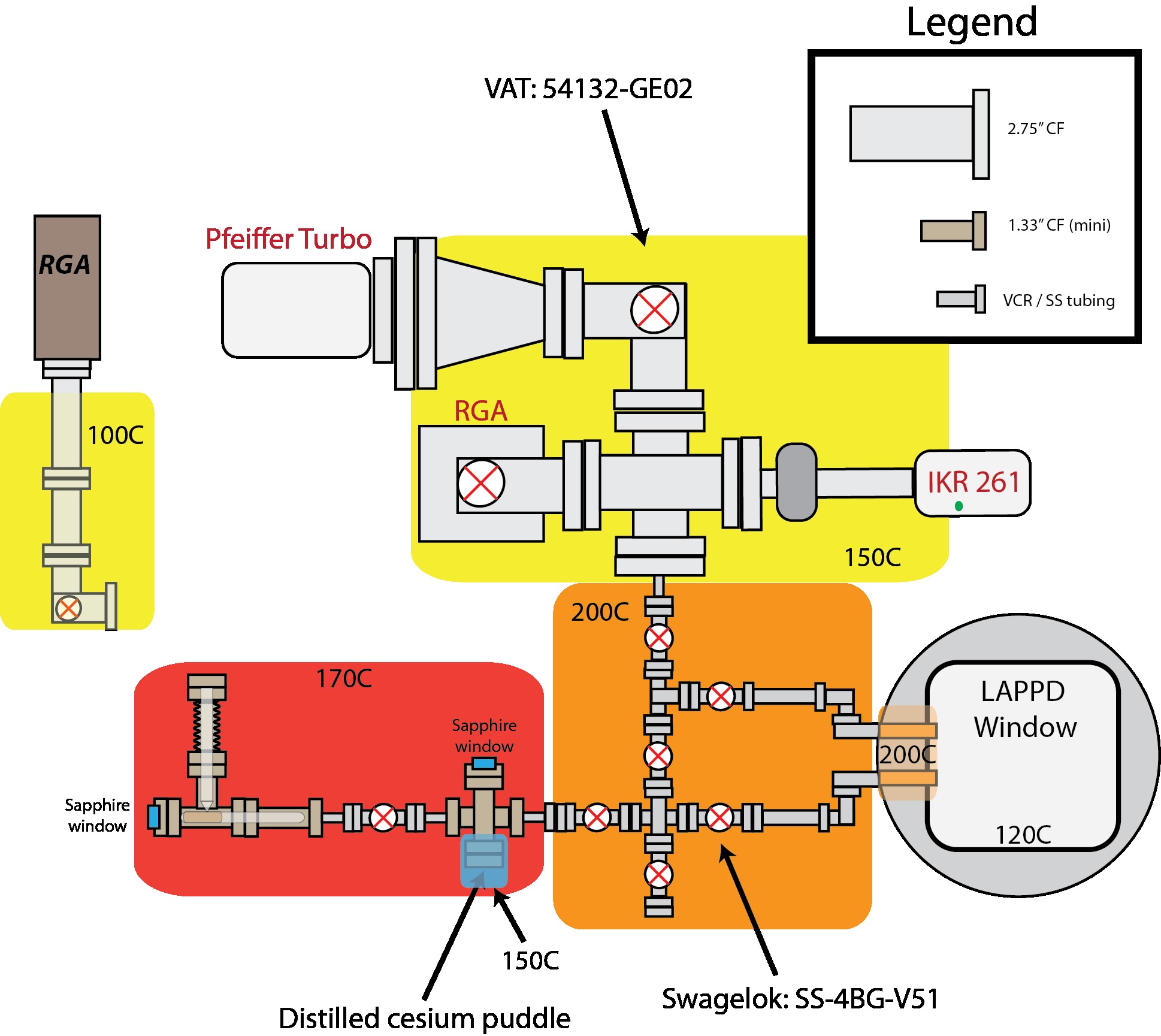} 
\caption{The physical layout and process temperatures during alkali vapor
  transport of the UHV inner vacuum manifold.}
\label{fig:manifold}
\end{figure}

The inner UHV vacuum system consists of the tile volume, copper
tubulations, and the manifold that connects the tile to
the inner system vacuum pump and the alkali vapor source.
A diagram of the manifold and the process temperatures during alkali transport
is shown in Figure~\ref{fig:manifold}.

%
%

%
%
\section{The \LAPPDTM Package}
\label{package}
\begin{figure}[htb]
\centering
\includegraphics[angle=0,width=0.48\textwidth]{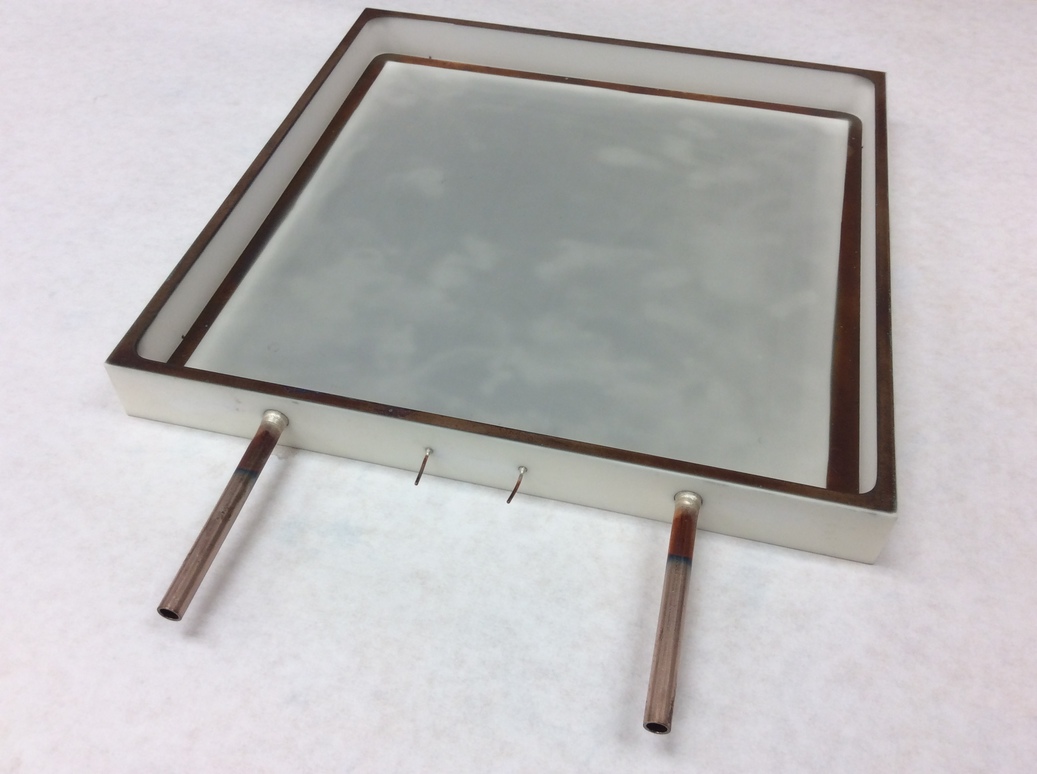}
\hfil
\includegraphics[angle=0,width=0.48\textwidth]{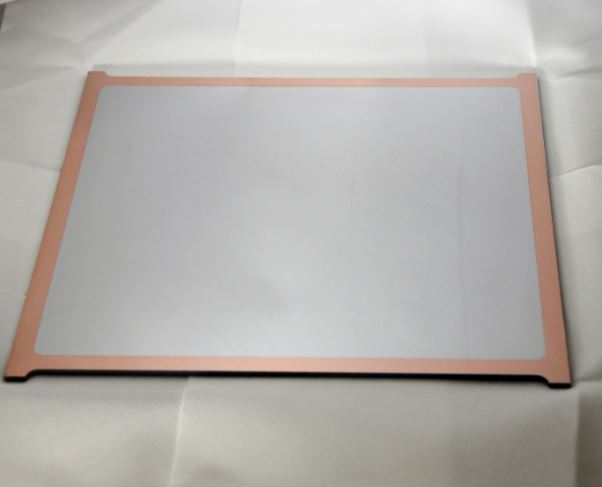}
\caption{Left: A two-pin ceramic tile body with metalized sealing surface and
  NiCr capacitively-coupled anode; Right: An entrance window with
  metalized sealing surface.}
\label{fig:body_and_window}
\end{figure}

The UC ceramic tile package consists of a monolithic ceramic tile base
with capacitively coupled anode and a transparent entrance window,
shown in Figure~\ref{fig:body_and_window}.  The fast signal pulses are
capacitively coupled through the anode on the bottom
plate~\cite{InsideOut_paper}, avoiding the difficulty in brazing a
large number of pins.


\begin{figure}[htb]
\centering
\includegraphics[angle=0,width=0.65\textwidth]{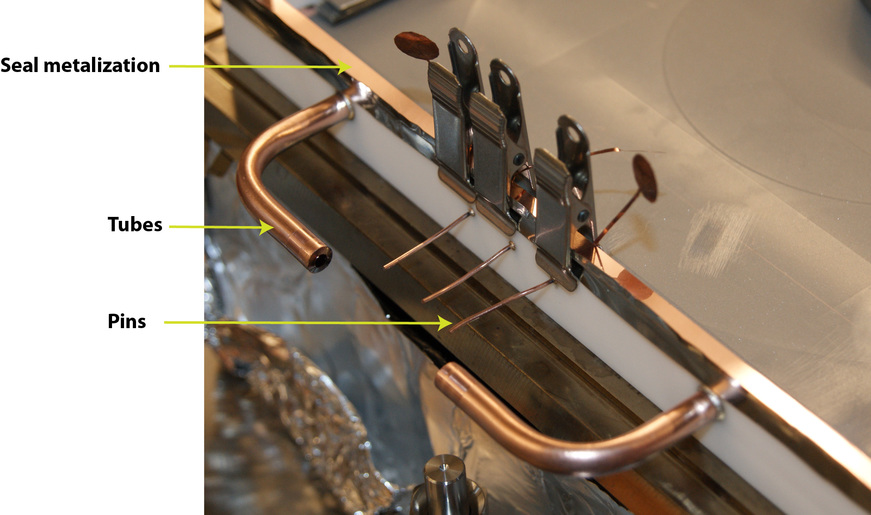} 
\caption{Copper tubes that connect the tile base to the UHV
  manifold, and copper pins that provide HV to the cathode and
  amplification stack.}
\label{fig:pins_and_tubes}
\end{figure}

The tile base is one-piece, composed of high-purity \Al2O3
ceramic~\cite{ceramic_vendors,Image_Library_304_315_316_320}. Two
1/4'' OFHC copper tubes are brazed through the sidewall to provide for
pump-down and alkali vapor transport. Current designs incorporate 6
copper pins, 3 per side, brazed through opposing sidewalls for HV
distribution to the cathode, amplification section, and
anode~\cite{Image_Library_304_315_316_320}. Figure~\ref{fig:pins_and_tubes}
shows a ceramic sidewall with pins and tubes.

The top surface of the tile base sidewall is the surface that mates
with the window to provide the hermetic seal. The solder seal requires
that this surface be metalized. The majority of the trials used thin
film evaporation to deposit 200 nm of 80-20 NiCr followed
by 200 nm of Cu without a vacuum
break~\cite{Image_Library_304_315_316_320} on the sealing surface,
with the sidewall and anode tile surfaces protected by masks.

The bottom interior surface of the tile was also coated by
evaporation. The first step was evaporation of 10 nm of NiCr on the
active area of the anode to provide capacitive coupling to the
external signal board~\cite{InsideOut_paper}. A second set of masks
was used to deposit 200 nm of NiCr followed by 200 nm of Cu, also
without a vacuum break, for a narrow border overlapping the 10 nm
NiCr anode to provide electrical
contact~\cite{Image_Library_304_315_316_320}.

The UHV inside the module is maintained after sealing by an
internal non-evaporative getter consisting of two stacks each containing
six 174-mm by 8-mm strips of SAES ST707~\cite{SAES_getter}
held in clips soldered to the anode.

The mating surface of the window was also metalized by
evaporation~\cite{Image_Library_304_315_316_320} using successive
masks~\cite{Clausing}. First 200 nm of NiCr was deposited on the
border with thin `fingers' toward the center to distribute current
across the cathode. A second evaporation deposited 200 nm of Cu on the
Ni border for the indium alloy solder bond. A third evaporation
deposited a thin Sb film as the pre-cursor for the alkali cathode.

The unsupported spans of the window and tile base anode
surface support almost 1000 lbs of compressive atmospheric
pressure.  Sixteen columns of 3/8''-diameter ceramic cylinders
(``buttons'') are placed internally at 1.8'' spacing in the gaps
between the window, MCP's, and anode. The buttons serve three
purposes:
\begin{enumerate}
\setlength{\itemsep}{-0.03in}
\item Supporting the package against the compressive force of
  atmospheric pressure on the top and bottom of the package;
\item Setting the gaps between the MCP's, tile base, and the photocathode;
\item Applying mechanical force between the MCP surfaces and the
  high-voltage internal electrodes that connect the accelerating high
  voltages to their respective pins through the sidewall.
\end{enumerate}

\begin{figure}[htb]
\centering
\includegraphics[angle=0,height=0.30\textwidth]{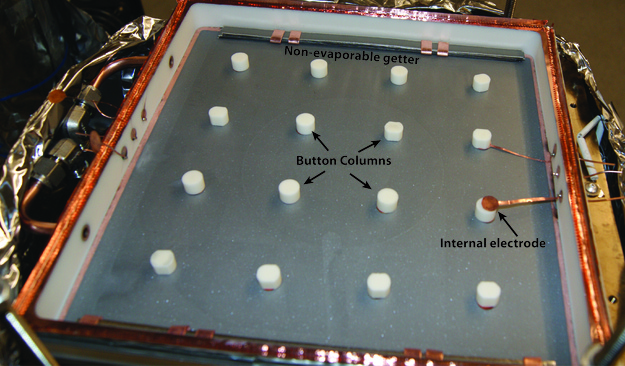} 
\hfil
\includegraphics[angle=0,height=0.30\textwidth]{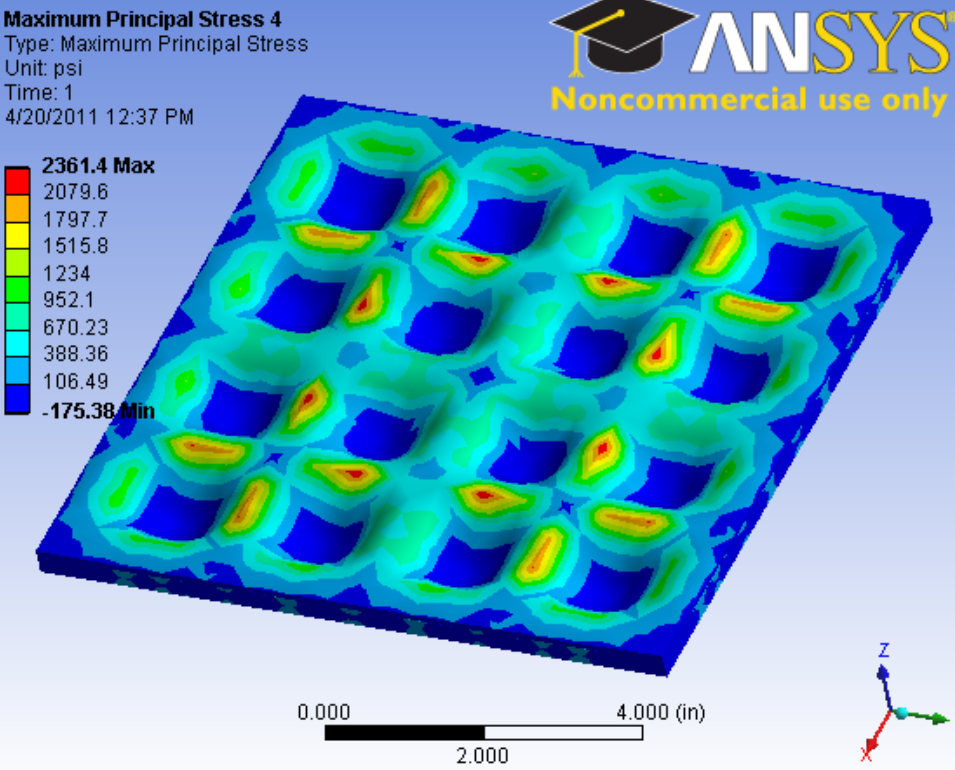} 
\caption{Left:The tile base during assembly showing the spacer buttons
  and their connecting internal high-voltage electrodes. At the top of
  the picture the getter is positioned to be soldered to the
  NiCr anode layer during the bake-out thermal
  cycle. Right: A finite-element analysis of the stress on
  a B33 window under vacuum.}
\label{fig:buttons_and_getters}
\end{figure}

The left-hand panel of Figure~\ref{fig:buttons_and_getters} shows the
placement of the button columns and the internal high-voltage
electrodes connected to the side-wall pins during the first stage of
assembly. Atmospheric pressure is used to constrain the amplification
stack against lateral motion; the right-hand panel shows the
predicted stress for a 2.75 mm-thick B33 glass window. 

The amplification section consists of two ALD-coated 20 cm-square MCPs
in a chevron configuration~\cite{Incom,Incom_production,Arradiance}.
The spacing between plates is set by button spacers to 0.080''; the
spacing between the top or bottom plate and the top (cathode) or
bottom (anode) surfaces, respectively, is similarly set to 0.25''. The
capillary seal (Section~\ref{seal}) reduces the necessary assembly
tolerances by precisely determining the indium layer thickness via a `hard-stop'
in the seal gap rather than by the stack height. The measured
precision on the column heights is 0.5 mils.  Before assembling the
tile the micro-channel plates were vacuum-baked and scrubbed with a UV
lamp.


The monolithic ceramic tile avoids the problem of having many
penetrations to bring the
signals out by capacitively coupling through the thin metal anode to an
external signal pickup board~\cite{InsideOut_paper}. The signal board
is customized to  use micro-strips~\cite{Tang_Naxos, RSI_paper} or
pads~\cite{InsideOut_paper}, determined by the application.

%
%

\section{The Hermetic Window Indium-Alloy Seal}
\label{seal}

The hermetic seal is one of the challenges of scaling up MCP-PMTs in
size. A solution consists of a clamping the window in place on the
tile base at a set gap width, typically 0.002'', and `wicking' molten
indium alloy\cite{InAg_alloy} into the gap by capillary action from
wire placed contiguous to the gap~\cite{Seal_patent}.  The sealing
surfaces require metalization prior to assembly.  Both sealing
surfaces are coated by evaporation with 200 nm of NiCr followed by 200
nm of Cu.

\begin{figure}[htb]
\centering
\includegraphics[angle=0,width=0.65\textwidth]{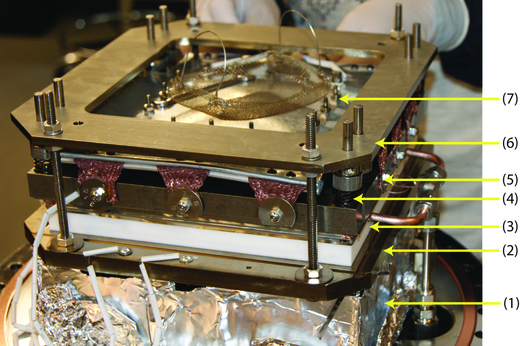} 
\caption{An assembled tile clamped in place in the compression fixture
  prior to sealing. Heaters above and below the tile couple to the top
  and bottom fixture plates.  The indicated components are: (1) the
  lower NiCr heater assembly; (2) the bottom compression fixture
  plate; (3) the tile assembly; (4) the compression mechanism; (5)
  floating rigid press bars; (6) the upper compression fixture plate;
  and (7) the upper NiCr heater assembly.}
\label{fig:tile_assembly}
\end{figure}

The tile base, inner amplification components, and window are clamped
in place in a mechanical `fixture' prior to the thermal sealing
cycle. The fixture consists of a 0.50''-thick SS bottom plate that
supports the tile base, a compression fixture to clamp the window onto
the base from above, heating elements, and thermocouples for measuring
the temperature during the process cycle.  The copper tubulations are
connected to the manifold to complete the inner vacuum
system. Figure~\ref{fig:tile_assembly} shows an assembled tile in the
fixture prior to sealing.

 To bake out the tile and form the hermetic seal, two heating
units, shown in Figure~\ref{fig:tile_assembly}, consisting of 3mm-thick
6''-square ceramic plates wound with NiCr heating
ribbon~\cite{heater_note} are placed in thermal contact with the lower and
and upper compression plates, as shown in
Figure~\ref{fig:tile_assembly}.  The outer vacuum vessel is not
heated. 

\begin{figure}[htb]
\centering
\includegraphics[angle=0,width=0.65\textwidth]{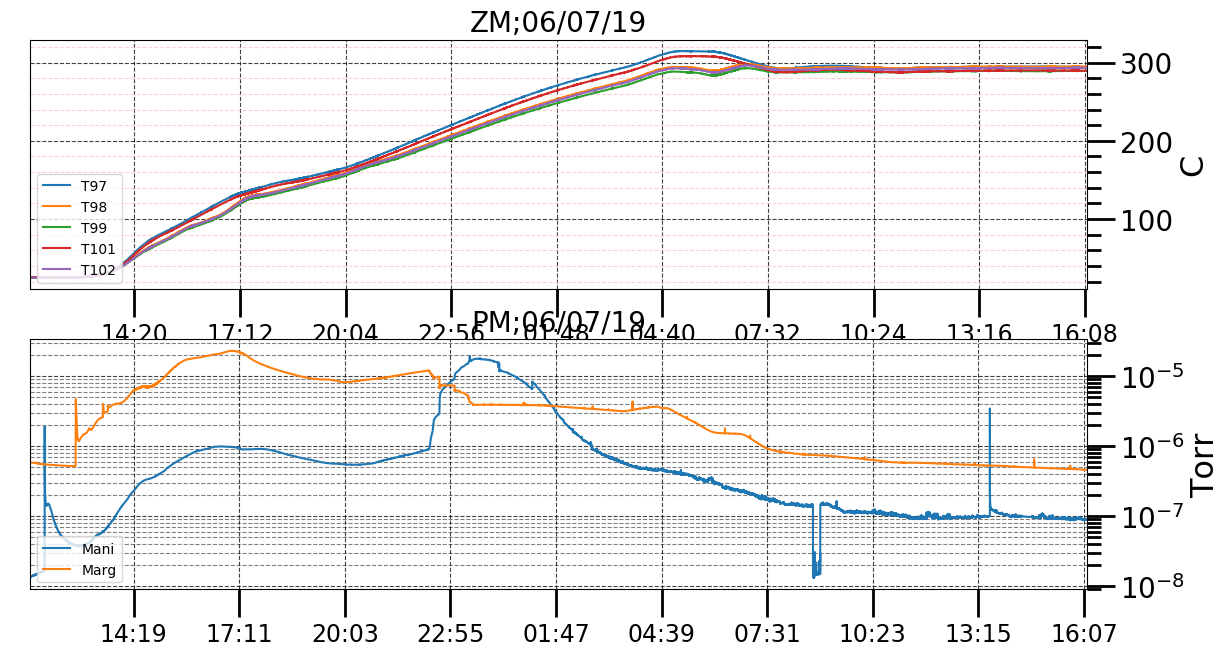}
\caption{The thermal cycle temperature and dual vacuum pressure data
  from Tile 31. The feature around 220~\degreesC is due to the
  formation of the capillary seal between the two vacuum systems after
  the In-Ag alloy has melted at a temperature around 143~\degreesC.}
\label{fig:tile29_thermalcycle}
\end{figure}

Figure~\ref{fig:tile29_thermalcycle} shows the thermal cycle for Tile
31. The cycle consists of a ramp up of approximately 15 hours, a
bake-out of 18 hours at 290~\degreesC, followed by cool-down to room
temperature. The feature around 220~\degreesC is due to the formation
of the indium alloy capillary seal.

%
Leak checking after sealing is performed in two steps. A global check
is performed before opening the outer vacuum vessel using a Residual
Gas Analyzer (RGA)~\cite{SRS_RGA} by back-filling the vessel with
argon with the tile isolated from the UHV (inner) manifold.  After
several hours, the pump to the manifold is valved off, and the valve
between the manifold and the tile is re-opened. The RGA can then
detect through the manifold any argon accumulated in the tile. We
estimate the sensitivity of the global check to be 1-2 $10^{-13}$
mbar-liters/sec.

If a leak is detected, after the outer vessel is opened a systematic
local search is performed with a helium leak
detector~\cite{helium_leak_detector} attached to the manifold and a
needle probe helium source. We estimate the sensitivity
of this local check to be several $10^{-11}$ mbar-liters/sec.

%
%
\section{Barois Photo-cathode Synthesis}
\label{photocathode}
The air-transfer assembly process involves depositing a Sb pre-cursor
layer on the window prior to assembly. After the sealing cycle is
complete and the tile is leak-tested, the tile is re-heated to a
temperature of $\approx$120~\degreesC, below the melting point of the
solder seal, for photocathode synthesis by introducing alkali
vapor~\cite{barois_paper_1,barois_paper_2}.


The bialkali \K2CsSb photocathode cathode compound was synthesized
directly as part of this
program~\cite{RMD_SBIR_1,RMD_SBIR_2,Cathode_patent}. In the process
development described here only a single alkali, Cs, was used for
simplicity.

\begin{figure}[htb]
\centering
\includegraphics[angle=0,width=0.65\textwidth]{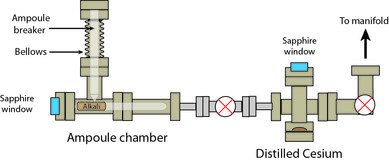} 
\caption{Diagram of the alkali (Cesium) vapor source. The left-hand
  section is the ampoule chamber; the right-hand section is the source
  chamber.}
\label{fig:cs_source}
\end{figure}

The cesium vapor source is shown schematically in
Figure~\ref{fig:cs_source}. A glass ampoule of
Cs~\cite{alfa_aesar,espi_metals} is inserted in the vertical bellows
section of the left-hand (ampoule) chamber.  The ampoule is broken
from above by a ceramic rod attached to a compressing the bellows.
The Cs liquid and glass ampoule can be observed through a sapphire
window.  The liquid Cs is then evaporated away from the glass shards
of the ampoule into the right-hand (source) chamber by heating the
ampoule chamber~\cite{Cs_transport_temps}.  The valve between the
source chamber and the ampoule chamber is then closed. A sapphire
window in the source chamber allows visual monitoring of the surface
of the distilled Cs. Cesiation of the tile is initiated by opening the
valve to the manifold with the source chamber at 150-155~\degreesC.

\begin{figure}[htb]
\centering
\includegraphics[angle=0,width=0.65\textwidth]{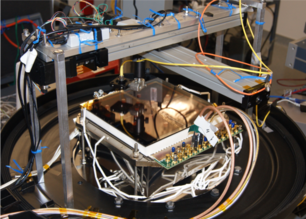} 
\caption{A 2D stage for mapping photo-response during and after
  photocathode synthesis. } 
\label{fig:2d_stage}
\end{figure}

A 2-dimensional motion stage for mapping the photocathode response
during photocathode synthesis is shown in
Figure~\ref{fig:2d_stage}. The module is still connected to the UHV
manifold; pinch-off occurs only after leak-checking and cathode
characterization.


\begin{figure}[htb]
\centering
\includegraphics[angle=0,width=0.65\textwidth]{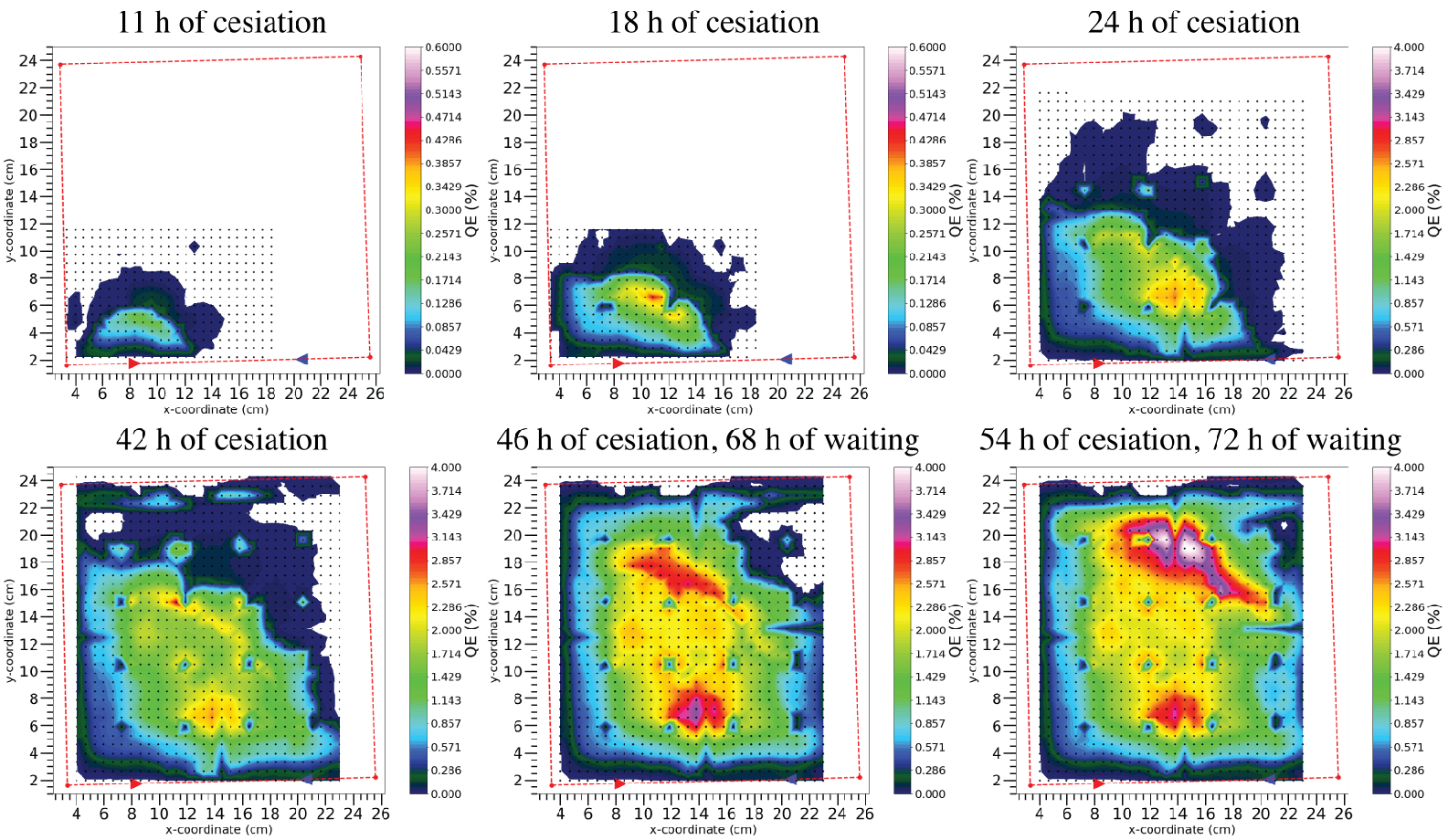} 
\caption{Successive maps of the photo-response for Tile 31 during
  photocathode synthesis.}
\label{fig:tile31_pc}
\end{figure}

Figure~\ref{fig:tile31_pc} shows successive maps of the photo-response
during the activation of the pre-cursor Sb layer by Cs vapor. The
response grows to a saturation value near the inlet and then continues to
the far side of the 8.26''-square window. The
response saturates near 3\% quantum efficiency (QE).

Neither the uniformity nor the end-point QE are satisfactory for a
useable commercial device.  We attribute the lower quantum efficiency
near the edges to a non-uniform temperature of the window during
synthesis. The tile was heated through its contact with the fixture
lower plate, and was subject to convective cooling from above.
Thermally isolating the assembly from convection should eliminate the
non-uniform temperature of the window. In addition, the QE saturated,
but at a lower value than is attainable for a good \Cs3Sb
cathode~\cite{Luca}. We attribute this to the overly-thick Sb
precursor layer.

\begin{figure}[htb]
\centering
\includegraphics[angle=0,width=0.65\textwidth]{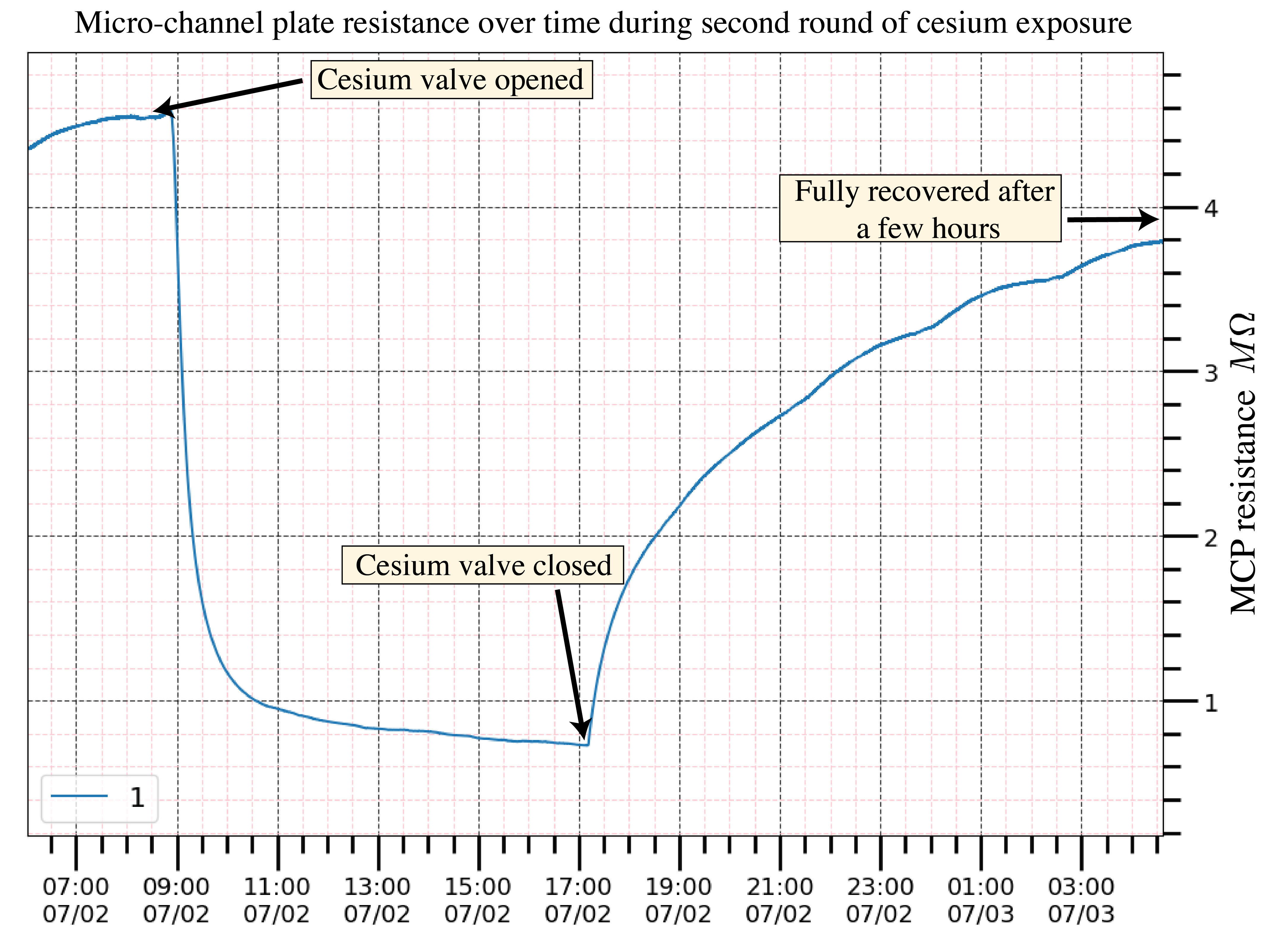} 
\caption{The resistance of the stack of two microchannel plates and
  spacers during a cesiation cycle, showing that the plates recover
  functionality after an initial large drop in resistance. } 
\label{fig:MCP_recovery}
\end{figure}

Figure~\ref{fig:MCP_recovery} shows that the micro-channel plates
recover functionality after cesiation. There is an initial large drop
in resistance that would preclude operation. After the alkali vapor is
valved off the plates increase in resistance and are operational after
cooling. The plates initially have increased gain and noise, but
the gain and noise fall back close to initial values within 
hours of operation.


%
%

After photocathode synthesis and characterization, the copper
tubulations are hermetically sealed by pinching
off~\cite{pinchoff_tool}. The razor-sharp edge of the pinch is then
potted for safety, and the tile can be removed from the fixture.

\begin{figure}[th]
\centering
\includegraphics[angle=0,width=0.50\textwidth]{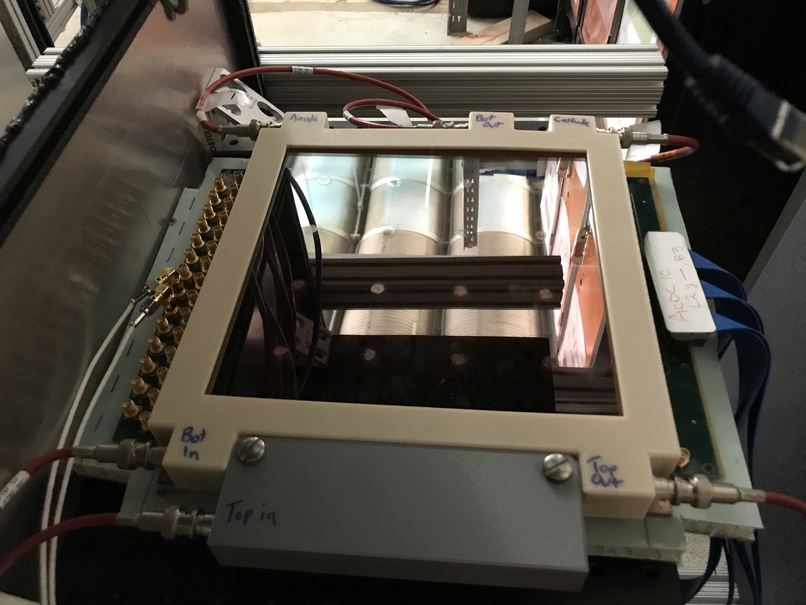} 
\caption{The UC \LAPPDTM Tile 31 with a signal readout board and HV
  connections. }
\label{fig:completed_lappd}
\end{figure}

%
%
\section{Conclusions}
\label{conclusions}

Large-area MCP-based photomultipliers employing ALD coated plates have
unique performance parameters, with typical gains greater than $10^7$,
amplification-section (as opposed to photocathode) noise less than a
few Hz, sub-mm resolutions in each transverse spatial dimension, and
time resolution of 5-10 psec for charged particles and 30-50 psec for
single photons~\cite{RSI_paper,timing_paper}. The small feature size,
concomitant good resolutions, and high gain are intrinsic rather than
highly tuned~\cite{Philadelphia_2017_proceedings}. The ALD-based
functionalization of the glass capillary
substrates~\cite{Incom,Incom_production} eliminates the shortened tube
lifetime due to ion feedback~\cite{LAPPD_proposal,Lehman}.

At the outset of this work the major obstacles to PMT-like batch production of
large-area MCP-PMTs were identified as:
a) the long and rectangular hermetic seal;
b) `in-situ' synthesis of alkali photo-cathodes 
in a large-area low-profile package;
c) compromised performance of the MCPs after exposure to
  alkali vapors; and
d) a robust package that supports large numbers of high bandwidth electrical
signal connections to an external readout system. We address these in
turn below.

The demonstrated solution to the hermetic seal relies on: a) clamping
the window at the pre-set width of the solder seal layer prior to
heating; clean Au or Cu surfaces on the window and base with no solder
material (and hence no oxides) in the gap prior to sealing; b)
allowing capillary action to 'wick' the molten alloy into the
window-base gap during bakeout away from oxide; and c)) use of In-Ag
alloy as solder.

After this work started, we discovered that the chemistry of the
'in-situ' photo-cathode synthesis starting with a pre-deposited
uniform thickness Sb layer was studied in depth 30 years
ago~\cite{barois_paper_1,barois_paper_2}. We also discovered that the
air transfer of the window with a pre-deposited Sb layer is currently
in use commercially~\cite{MELZ}. We note that
Ref.~\cite{barois_paper_2} states ``Manufacturing photo-emissive
layers by equilibrating the antimonides with binary alkali vapors
would lead to a product with a better definition than those obtained
through the dynamic process used at present''.

The recovery of MgO-coated MCP functionality after exposure to Cs
vapor during photocathode synthesis was demonstrated. The plates
initially drop dramatically in resistance, and both gain and noise
increased,  but recover with subsequent operation before pinch-off.

Lastly, the development of capacitive coupling of an internal thin
metal anode layer to an external, application-specific, signal-pickup
printed circuit board enables the decoupling of the photomodule design
from the optimization of pads or strips for high-bandwidth signals for
applications requiring many pads or strips in the
readout~\cite{InsideOut_paper}.

We have shown that each of these potential `show-stopper' obstacles is
technical rather than fundamental. A proof-of principle solution to
each that is scalable to batch production is described above. However,
the development of an optimized industrial batch process for bringing
volume up and cost down for wide-spread use in particle physics,
medical imaging, and nuclear security remains as the task for
commercialization.

\section{Acknowledgements}
\label{acknowledgements}

The University of Chicago is supported by the HEP Division of the
Department of Energy under con tract DE-SC-0008172 and DE-SC-0020078,
by the National Science Foundation under grant PHY-106601 4, and by
the Physical Sciences Division of the University.  The development at
Incom is supported by the DOE Nuclear Physics Division through award
number DE-SC0015267.

We thank the many collaborators and experts who have
contributed. Special thanks for critical interventions are due to
Oswald Siegmund, Anton Tremsin, Jason McPhate, Howard Nicholson, Klaus
Attenkofer, Sergey Belyanchenko (MELZ), Howard Clausing (Clausing
LLC), Jeffrey DeFazio (Photonis), Igor Fedorov (MELZ), Mary Heintz, Giles
Humpston, Robert Jarrett (Indium Corp.), Christine Mariano
(International Ceramics Engineering), Helmut Marsiske, Rich Northrop,
Matthew Poelker, Charles Sinclair, and Neal Sullivan (Arradiance).


\begin{thebibliography}{99}
\bibitem{Cs_transport_temps} During distillation the ampoule chamber is
  heated to 240~\degreesC and the source chamber to 210~\degreesC.
\end{thebibliography}
\end{document}